# Biomaterials: A trendy source to engineer functional entities – An overview

Hafiz M. N. Iqbal*

Tecnologico de Monterrey, School of Engineering and Sciences, Campus Monterrey, Ave. Eugenio Garza Sada 2501, Monterrey, N. L., CP 64849, Mexico. *Corresponding author: Tel: +52 (81) 83582000 ext. 5679. E-mail: hafiz.iqbal@itesm.mx.

**Abstract**
The biomaterials exploitation in a sophisticated manner can provide extensive opportunities for experimentation in the field of interdisciplinary and multidisciplinary scientific research. Owing to the unique features of this trendy area, research scientists have been directed/redirected their interests in bio-based biomaterials for targeted applications in different sectors of the modern world. The present manuscript highlights the novel perspectives of biomaterials as a trendy source to engineer functional entities in numerous geometries for pharmaceuticals, cosmeceuticals, nutraceuticals, and other biotechnological or biomedical applications.

**Introduction**
Biomaterials have gained special research interests as novel candidates and potential alternatives to the traditional counterparts [1-3]. In the past few years, numerous efforts have been made to engineer new types of added-value products e.g., bio-composites, antibacterial active materials with natural phenols as functional entities, materials-based carriers for drug delivery purposes etc. [3-13]. This trendy area is now moving towards the development of 'greener' technologies and in turn, the principle of 'going green' has directed this search towards eco-friendly, sustainable and functional materials. Words like renewability, recyclability, and sustainability are emphasized in growing environmental awareness. The fact is that environmental legislation is the driving force behind the development of these materials [2]. With ever increasing scientific research, knowledge and socioeconomic awareness, industrial communities are now more concerned about the environmental impact of persistent plastic-based wastes and their synthetic counterparts. The divergence from synthetic to biomaterials is becoming the center of interest for scientific communities and research-based organizations, around the globe. Therefore, in this context, bio-based materials are being engineered for target applications in different sectors of the modern world.
Research is underway around the world on the development of 'greener' polymer technologies. In recent decades, there has been a growing search for new high-performance products for multipurpose applications in biotechnology at large and biomedical, pharmaceutical and/or cosmeceutical in particular [14]. The principle of 'going green' has directed this search towards eco-friendly materials with multifunctional features [2, 11]. One area that has received limited attention so far, but that will gain in importance as naturally conferring antimicrobial agents use becomes further established, is the incorporation of such novel agents into the materials to provide an antibacterial effect on contact of that material with the target bacterium. Such antimicrobial active biomaterials might have great potential to respond to a new infection before the clinical signs are evident, with the potential to significantly improve patient prognosis. Antimicrobial agents-impregnated materials could be used as medical implants and in applications relevant to hospital hygiene. However, there are also clear industrial and biotechnological requests for materials that are loaded with natural agents that can quickly prevent deleterious microbial action following contamination events. It is intended that a technology platform for future exploitation, e.g. *in vivo* and *ex vivo* designs to find out other suitable potential applications such as biomedical implants of these newly developed novel materials, could also be established.

**Potential applications of biomaterials**
The biomaterials exploitation for biotechnological applications at large and biomedical in particular has several intrinsic advantages that include but not limited to the biocompatibility, biodegradability, renewability, sustainability, and non-toxicity [2, 15]. The sustainability concept is shown in Figure 1. From the application view point, a wider spectrum of biomaterials and biomaterials based novel constructs has been engineered for target applications with a particular reference to the active antimicrobial constructs. Some of the major examples include bacterial cellulose (BC), collagen, PLA, and chitosan. All these materials have been characterized and well organized/developed into value-added structures, thus can provide a proper route to emulate bio-systems - a biomimetic approach.

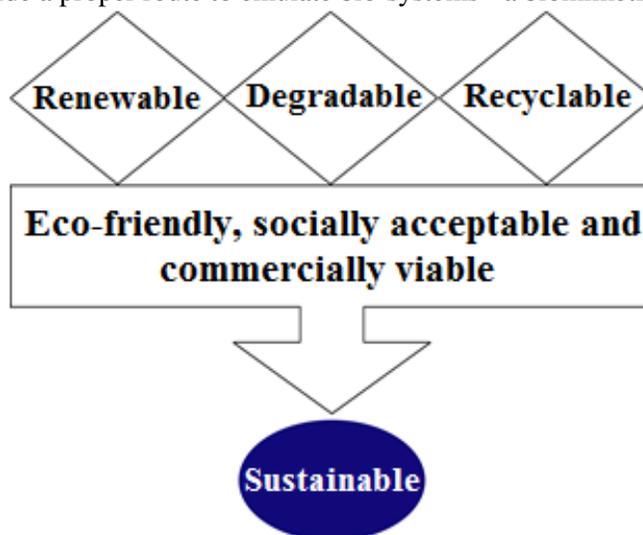

**Figure 1** Concept of "sustainability" (Reproduced with permission from Ref. [2].

**Bacterial Cellulose (BC) – Novel aspects**
Apart from plants, cellulose is biosynthesized by certain bacteria, *e.g.*, *Rhizobium* spp., *Agrobacterium* spp., *Acetobacter* spp., and *Alcaligenes* spp. [2, 16] and termed either bacterial cellulose (BC). Figure 2 illustrates an overview of the BC network produced by bacteria (*Acetobacter xylinum*). BC is a straight chain polysaccharide, with the same chemical structure as cellulose that is derived from plants. BC has notable advantage of being devoid of lignin, pectin, hemicellulose and other biogenic products that are normally associated with plant cell wall structures [6, 17]. Because of its high purity and special physicochemical characteristics, BC has applications in a wide range of sectors, including food, bio-medical (*e.g.*, wound care), and tissue engineering (*e.g.*, nanocomposites) [2, 6, 15, 18-21]. Therefore, in light of the afore-mentioned characteristics, BC may be a promising candidate for the development of value-added products.

Among the possible alternatives, the development of composites, utilizing cellulose as a reinforcement material, are under investigation in almost every industry. There are various methods of manufacturing bio-based products, depending on the processing techniques; *e.g.*, surface casting, ultrasonic-assisted casting, pultrusion, extrusion, injection molding, press molding, hand lay-up, filament winding, sheet molding compounding, and enzymatic grafting [6-8, 22, 23]. Moreover, cellulose offers the ability for surface modification, eco-friendly processing, non-toxic nature, easy handling, and no health risks, while most synthetic polymers pose significant health risks, such as skin irritation and respiratory disease [6, 24]. Cellulose and cellulose-based materials can be used for different applications, including food, paper and packaging, tissue engineering, pharmaceutical, cosmeceutical, electronics, dentistry, and medicine [25-27]. Figure 3 illustrates various biomedical applications of bacterial cellulose (BC) as a model example from bio-based biomaterials.

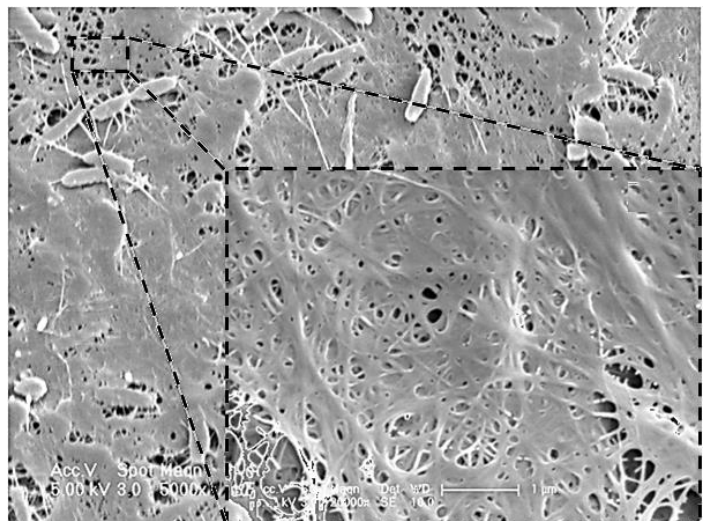

**Fig. 2.** Scanning electron microscope image of a bacteria-generated bacterial cellulose network. Reproduced with permission from Ref. [2].

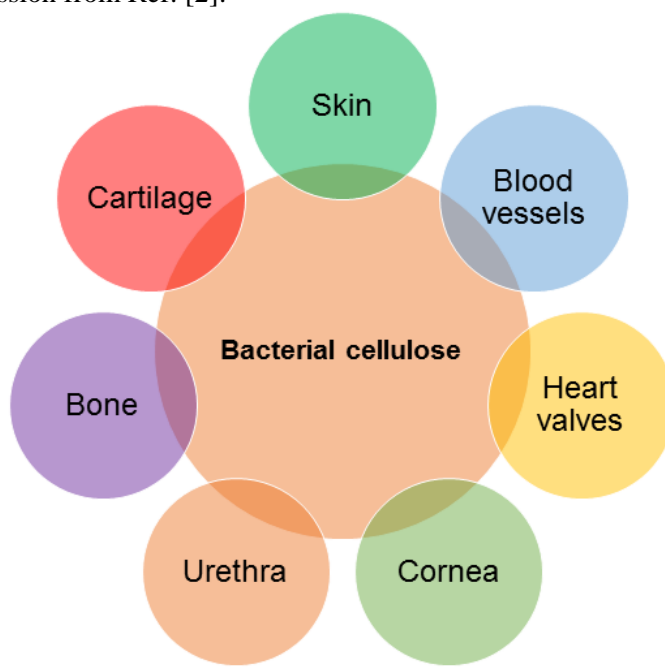

**Figure 3** Prospects for the various biomedical applications of BC and BC-based materials. (Reproduced with permission from Ref. [15]).

**Potential Applications of BC**
The high level of purity, change in color, change in flavor, and enormous potential to develop various shapes and textures, makes BC/MC a potential candidate for the food industry. The most popular use of BC in food is the production of Nata, originating from the Philippines; Nata is a traditional sweet dessert in Southeast Asia. Nata is a fermentation product of the bacteria, *Acetobacter xylinum*. Referred to as Nata de coco and Nata de pina, their flavors are controlled by the coconut water-based and pineapple water-based culture mediums, respectively [28, 29]. Figures 4 and 5 illustrate the schematic representation of the production process for Nata de coco and Nata de pina, respectively [15].

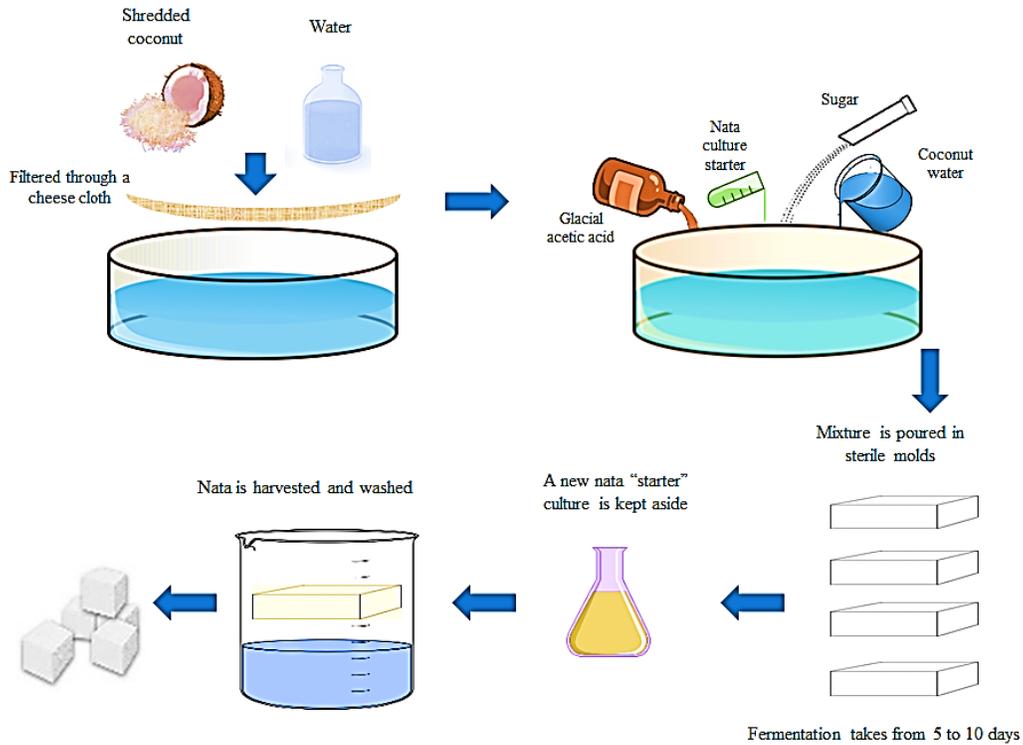

**Fig. 4.** Schematic representation of the Nata de coco production process. (Reproduced with permission from Ref. [15]).

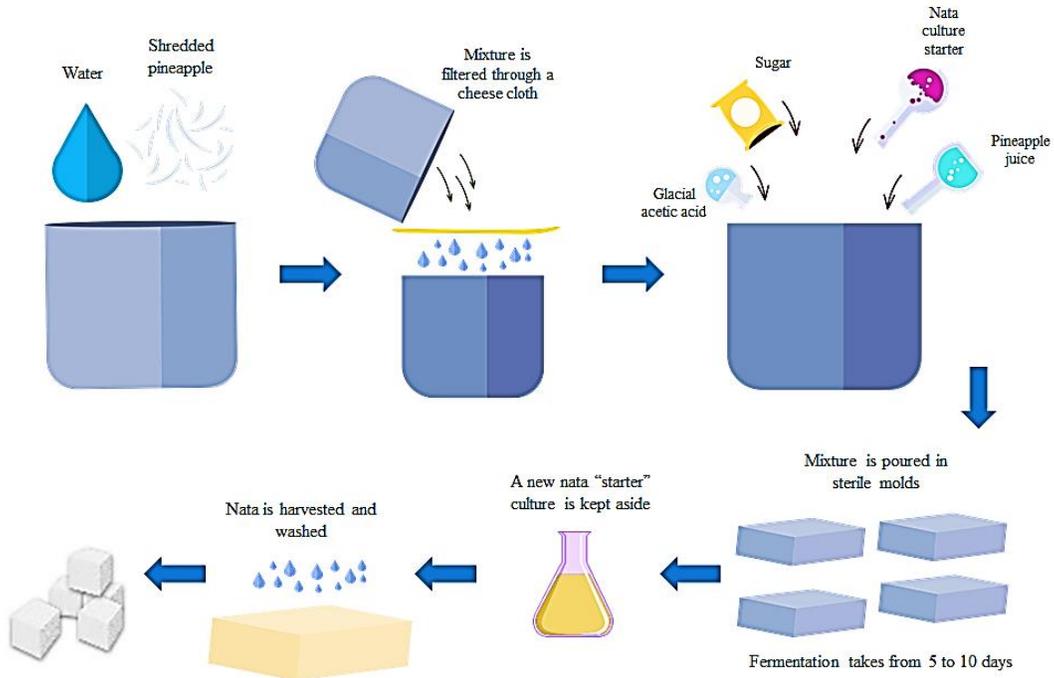

**Fig. 5.** Schematic representation of the Nata de pina production process. (Reproduced with permission from Ref. [15]).

**Antimicrobial active materials**
Bio-based biomaterials are moving into the mainstream applications changing the dynamics of 21st-century materials and their utilization in drug delivery strategies. Owing to the increasing consciousness and demands to reduce bacterial contaminations in healthcare facilities and possibly to cut pathogenic infections, the engineering aspects of novel active anti-microbial materials are considered to be a potential solution to such a problematic issue [2]. These materials have not only been a motivating factor for the materials scientists, but also they provide potential opportunities for improving the living standard [2, 30]. In past years, several authors have already been reported antibacterial features of several materials including silver nanoparticles [31-35]. However, excess release of silver nanoparticles inhibits osteoblasts growth and can also cause many severe side effects such as cytotoxicity [31] (Wang et al., 2014). Therefore, there is a persistent need to prepare green composites using one or more individual biopolymers to reduce or even eliminate the risk of bacterial infection without impairing the cytotoxicity capabilities. The antibacterial mechanism of natural phenols is naturally concomitant due to the presence of active hydroxyl groups. This is because the interaction between natural phenols and bacteria can change the metabolic activity of bacteria and eventually cause their death [7-9]. Based on an earlier published data, most of the phenolic compounds including gallic acid, *p*-4-hydroxybenzoic acid, and thymol have an ability to disrupt the lipid structure of the bacterial cell wall, further leading to a destruction of the cell membrane, cytoplasmic leakage, and cell lysis which ultimately leads towards the cell death [2].

**Biomaterials based biocomposites**
As mentioned above, there has been increasing research interest in the development of biomaterials-based bio-composites with multi characteristics i.e. (1) stronger, (2) stiffer, (3) lighter along with other multi-functional properties for a variety of industrial and biotechnological applications [2]. A composite is defined as a "material that consists of two or more distinct materials/polymers in order to obtain tailor-made characteristics or to improve or impart ideal properties". More importantly, tailor-made characteristics include but not limited to the specific strength, thermal properties, surface properties, biocompatibility, and biodegradability features that the individual material fails to demonstrate on its own [10, 23]. Whereas, a biocomposite can be defined as "composite materials derived from a biological origin and comprise on one or more phases are termed as bio-composites" [22]. A broad definition of a bio-composite is a composite material made up of natural or bio-derived polymers, *e.g.*, BC, PHAs, and PLA [23]. So far, a range of methodologies have been successfully adopted for the production of BC and BC-based composites [2]. Furthermore, potential applications of BC and BC-based composites are also provided in Table 1. In recent years, cellulose-based materials have been widely employed in the area of infection free wound healing, tissue engineering/implants applications. Naturally occurring phenol grafted P(3HB)-EC bio-composites, on the other hand, are expected to exhibit an excellent HaCaT compatibility and potentially favorable for cell proliferation. Recently, Iqbal et al. [7] have developed a series of novel bio-composites with natural phenols as functional entities and P(3HB)-EC as a base material using laccase as a grafting tool. I*n-vitro* biocompatibility of CA-*g*-P(3HB)-EC composites i.e., 0CA-*g*-P(3HB)-EC (control composite); 5CA-*g*-P(3HB)-EC; 10CA-*g*-P(3HB)-EC; 15CA-*g*-P(3HB)-EC and 20CA-*g*-P(3HB)-EC was achieved with the human keratinocytes-like HaCaT cells [2, 8]. Additionally, the morphologies of cell cultured from all of the test composites displayed healthy shape at 5 days, nevertheless, the amount of HaCaT cells seeded on the surface of 15CA-*g*-P(3HB)-EC composite was higher than those of 20CA-*g*-P(3HB)-EC composite (Figure 6).

**Table 1** Potential/Proposed Applications of Some Bacterial Cellulose-based "Green" Composite Materials. (Reproduced with permission from Ref. [15]).

| BC-based Materials | Methodology | New/improved functionalities | Potential/Proposed Applications | References |
|---|---|---|---|---|
| BC/Chi/Alg | Molding | Physical, mechanical, Biocompatibility | Wound dressing | [36] |
| BC-Vaccarin | Immersion | Physical, mechanical, and biocompatibility | Wound dressing | [37] |
| BC–xGnP | Impregnation | Thermal properties and electrical conductivity | Biosensors, tissue engineering | [38] |
| BC-$Fe_2O_3$ | Immersion | Magnetic behavior | Magnetic paper, loudspeaker membranes | [39] |
| BC–HA | Immersion | Biocompatibility | Bone tissue regeneration | [40] |
| P(3HB)-$g$-BC | Laccase-assisted grafting | Thermo-mechanical strength | Bio-plastics, Biomedical | [6] |
| AMPS-$g$-BC | Ultraviolet-induced polymerization | Conductivity, effective methanol barrier | Fuel cells | [41] |
| BC-MMTs | Immersion | Antibacterial properties | Wound dressing, regeneration materials | [42] |
| BC/GO | Vacuum-assisted self-assembly | Thermal, mechanical, conducting properties | Biochemical and electrochemical devices | [43] |
| BC-PAni | Immersion | Electrical conductivity | Flexible electrodes, flexible display devices, bio-sensors etc. | [44] |
| BC-MMT | Impregnation | Physical and mechanical properties | Biomedical | [27] |
| PANI/BC | Oxidative polymerization | Thermal, mechanical, conductivity | Flexible electrodes, display, sensors | [45] |
| BC/Chi | Immersion | Physical, mechanical, Biocompatibility | Wound dressing | [46] |
| ε-PL/BC | Immersion | Physical, Antibacterial | Packaging | [47] |

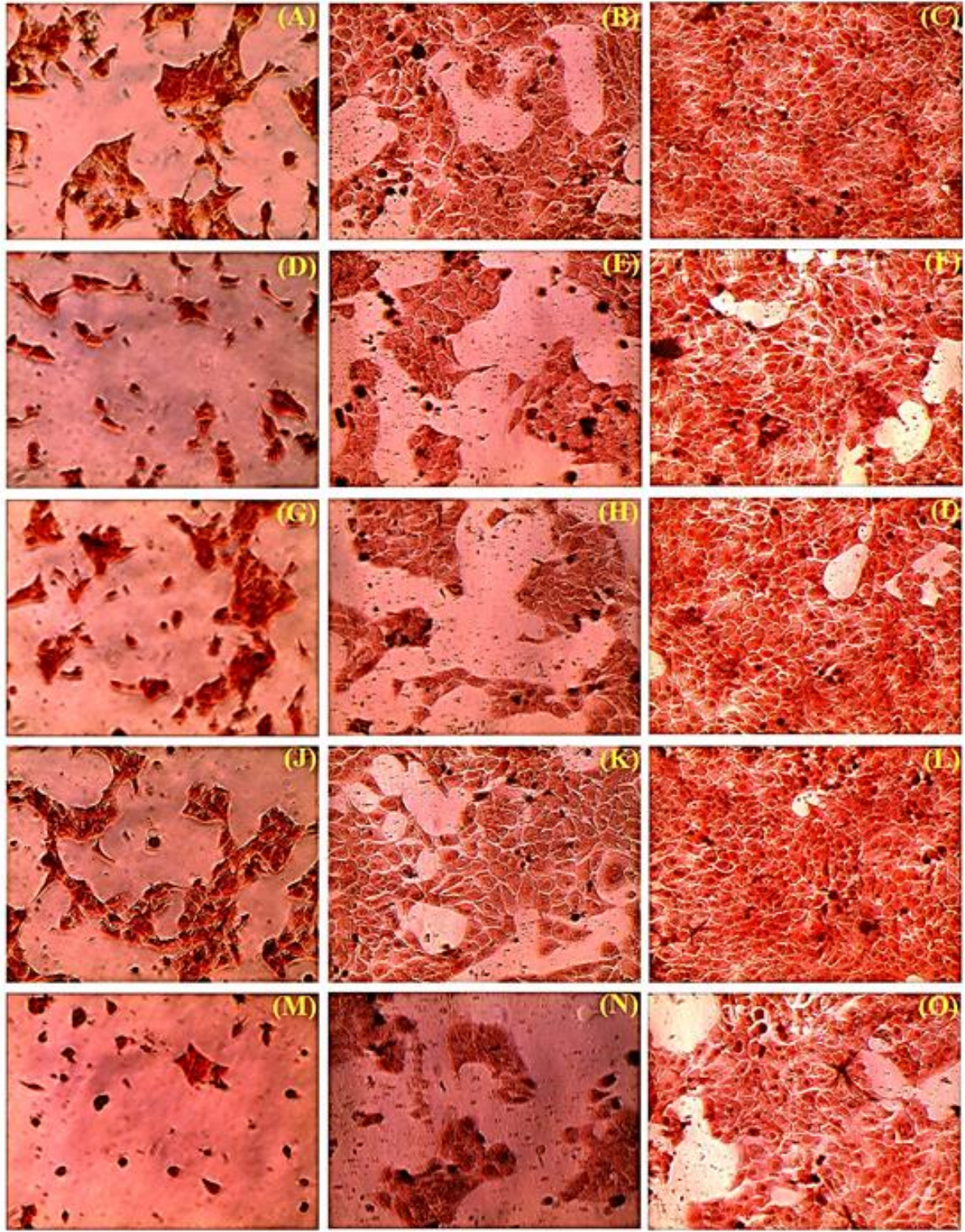

**Figure 6** Adherent morphology of stained images of human keratinocytes-like HaCaT cells seeded onto the composite surfaces. Images A, B, and C represent the HaCaT cells on native P(3HB)-EC composite (i.e., 0CA-*g*-P(3HB)-EC) after 1, 3 and 5 days of incubation, respectively; images D, E, and F represent the adhered HaCaT cells on 5CA-*g*-P(3HB)-EC composite after 1, 3 and 5 days of incubation,

respectively; images G, H and I represent the adhered HaCaT cells on 10CA-*g*-P(3HB)-EC composite after 1, 3 and 5 days of incubation, respectively; images J, K, and L, represent the adhered HaCaT cells on 15CA-*g*-P(3HB)-EC composite after 1, 3 and 5 days of incubation, respectively and images M, N and O represent the adhered HaCaT cells on 20CA-*g*-P(3HB)-EC composite after 1, 3 and 5 days of incubation, respectively. All of the test samples were stained using neutral red dye (5 mg/mL) for 1 h followed by three consecutive washings with PBS at an ambient temperature. All images were taken at 100X magnification (Reproduced with permission from Ref. [2, 11]).

**Concluding remarks and future considerations**
In conclusion, through sophisticated design and novel characteristics, the material can be modified to achieve an optimal infective capability. Such materials include but not limited to the biodegradable and biocompatible films and highly porous 3-D constructs. Bio-based biomaterials are versatile to synthesize novel constructs with multifunctional characteristics for potential applications in the biomedical sector. Moreover, a novel type of potent materials could be designed for the management and skin regeneration/repair from injury, particularly burns and ulcers, where the risk of bacterial infection is high. Material structure and performance integrity need to be accessed using a range of analytical and imaging techniques. Considering this scenario, research, production, and commercialization of such novel materials have drawn global efforts from numerous transnational companies as well as highly skilled research groups from around the world and diverse research areas.

**Conflict of interest**
Author declare no conflicting, competing and financial interests in any capacity.

**Acknowledgement**
This paper was supported by the Emerging Technologies Research Group and the Environmental Bioprocesses of Tecnologico de Monterrey, Mexico.

**References**
[1] Iqbal, H. M. N., Kyazze, G., & Keshavarz, T. (2013). Advances in the valorization of lignocellulosic materials by biotechnology: an overview. *BioResources*, *8*(2), 3157-3176.
[2] Iqbal, H. M. N. (2015). *Development of bio-composites with novel characteristics through enzymatic grafting* (Doctoral dissertation, University of Westminster).
[3] Bedian, L., Rodríguez, A. M. V., Vargas, G. H., Parra-Saldivar, R., & Iqbal, H. M. N. (2017). Bio-based materials with novel characteristics for tissue engineering applications–A review. *International Journal of Biological Macromolecules*, 98, 837-846.
[4] Iqbal, H. M. N., Kyazze, G., Tron, T., & Keshavarz, T. (2014a). "One-pot" synthesis and characterisation of novel P (3HB)–ethyl cellulose based graft composites through lipase catalysed esterification. *Polymer Chemistry*, 5(24), 7004-7012.
[5] Iqbal, H. M. N., Kyazze, G., Tron, T., & Keshavarz, T. (2014b). A preliminary study on the development and characterisation of enzymatically grafted P (3HB)-ethyl cellulose based novel composites. *Cellulose*, 21(5), 3613-3621.
[6] Iqbal, H. M. N., Kyazze, G., Tron, T., & Keshavarz, T. (2014c). Laccase-assisted grafting of poly (3-hydroxybutyrate) onto the bacterial cellulose as backbone polymer: Development and characterisation. *Carbohydrate polymers*, 113, 131-137.
[7] Iqbal, H. M. N., Kyazze, G., Locke, I. C., Tron, T., & Keshavarz, T. (2015a). In situ development of self-defensive antibacterial biomaterials: phenol-g-keratin-EC based bio-composites with characteristics for biomedical applications. *Green Chemistry*, *17*(7), 3858-3869.
[8] Iqbal, H. M. N., Kyazze, G., Locke, I. C., Tron, T., & Keshavarz, T. (2015b). Poly (3-hydroxybutyrate)-ethyl cellulose based bio-composites with novel characteristics for infection free wound healing application. *International journal of biological macromolecules*, *81*, 552-559.

[9] Iqbal, H. M. N., Kyazze, G., Locke, I. C., Tron, T., & Keshavarz, T. (2015c). Development of bio-composites with novel characteristics: Evaluation of phenol-induced antibacterial, biocompatible and biodegradable behaviours. *Carbohydrate polymers*, *131*, 197-207.

[10] Iqbal, H. M. N., Kyazze, G., Tron, T., & Keshavarz, T. (2015d). Laccase-assisted approach to graft multifunctional materials of interest: Keratin-EC based novel composites and their characterisation. *Macromolecular Materials and Engineering*, *300*(7), 712-720.

[11] Iqbal, H. M. N., Kyazze, G., Locke, I. C., Tron, T., & Keshavarz, T. (2015e). Development of novel antibacterial active, HaCaT biocompatible and biodegradable CA-gP (3HB)-EC biocomposites with caffeic acid as a functional entity. *Express Polymer Letters*, *9*, 764-772.

[12] Ahmad, Z., Asgher, M. & Iqbal, H. M.N. (2017). Enzyme-treated Wheat Straw-based PVOH Bio-composites: Development and characterization. BioResources, 12(2), 2830-2845.

[13] Asgher, M., Ahmad, Z., & Iqbal, H. M. N. (2017). Bacterial cellulose-assisted de-lignified wheat straw-PVA based bio-composites with novel characteristics. Carbohydrate Polymers, 161, 244-252.

[14] Ruiz-Ruiz, F., Mancera-Andrade, E.I., & Iqbal H.M.N., (2016). Marine-Derived Bioactive Peptides for Biomedical Sectors – A Review. *Protein & Peptide Letters, 24,* 109-117.

[15] Gallegos, A. M. A., Carrera, S. H., Parra, R., Keshavarz, T., & Iqbal, H. M. N. (2016). Bacterial Cellulose: A Sustainable Source to Develop Value-Added Products–A Review. *BioResources*, *11*(2), 5641-5655.

[16] Vandamme, E. J., De Baets, S., Vanbaelen, A., Joris, K., & De Wulf, P. (1998). Improved production of bacterial cellulose and its application potential. *Polymer Degradation and Stability,* 59(1-3), 93-99.

[17] Jonas, R., & Farah L. F. (1998). Production and application of microbial cellulose. *Polymer Degradation and Stability,* 59(1-3), 101-106.

[18] Svensson, A., Nicklasson, E., Harrah, T., Panilaitis, B., Kaplan, D. L., Brittberg, M., & Gatenholm, P. (2005). Bacterial cellulose as a potential scaffold for tissue engineering of cartilage. *Biomaterials,* 26(4), 419-431.

[19] Czaja, W., Krystynowicz, A., Bielecki, S., & Brown, R. M. (2006). Microbial cellulose - The natural power to heal wounds. *Biomaterials,* 27(2), 145-151.

[20] Shah, N., Ul-Islam, M., Khattak, W. A., & Park, J. K. (2013). Overview of bacterial cellulose composites: A multipurpose advanced material. *Carbohydrate Polymers,* 98(2), 1585-1598.

[21] Silva, N. C., Miranda, J. S., Bolina, I. C., Silva, W. C., Hirata, D. B., de Castro, H. F., & Mendes, A. A. (2014). Immobilization of porcine pancreatic lipase on poly-hydroxybutyrate particles for the production of ethyl esters from macaw palm oils and pineapple flavor. *Biochemical Engineering Journal,* 82, 139-149.

[22] Fowler, P. A., Hughes, J. M., & Elias, R. M. (2006). Biocomposites: Technology, environmental credentials and market forces. *Journal of the Science of Food & Agriculture,* 86(12), 1781-1789.

[23] Iqbal, H. M. N., Kyazze, G., Tron, T., & Keshavarz, T. (2016). Laccase from Aspergillus niger: A novel tool to graft multifunctional materials of interests and their characterization. *Saudi Journal of Biological Sciences*, In-Press. DOI: 10.1016/j.sjbs.2016.01.027.

[24] Yang, H.-S., Kim, H.-J., Son, J., Park, H.-J., Lee, B.-J., & Hwang, T.-S. (2004). Rice-husk flour filled polypropylene composites: Mechanical and morphological study. *Composite Structures,* 63(3-4), 305-312.

[25] Wang, Y., & Chen, L. (2011). Impacts of nanowhisker on formation kinetics and properties of all-cellulose composite gels. *Carbohydrate Polymers,* 83(4), 1937-1946.

[26] Mathew, A. P., Oksman, K., Pierron, D., & Harmand, M. F. (2012). Fibrous cellulose nanocomposite scaffolds prepared by partial dissolution for potential use as ligament or tendon substitutes. *Carbohydrate Polymers,* 87(3), 2291-2298.

[27] Ul-Islam, M., Khan, T., & Park, J. K. (2012). Water holding and release properties of bacterial cellulose obtained by *in situ* and *ex situ* modification. *Carbohydrate Polymers,* 88(2), 596-603.

[28] Shi, Z., Zhang, Y., Phillips, G. O., & Yang, G. (2014). Utilization of bacterial cellulose in food. *Food Hydrocolloids,* 35, 539-545.

[29] Jozala, A. F., de Lencastre-Novaes, L. C., Lopes, A. M., de Carvalho Santos-Ebinuma, V., Mazzola, P. G., Pessoa-Jr, A., Grotto, D., Gerenutti, M., & Chaud, M. V. (2016). Bacterial nanocellulose production and application: a 10-year overview. *Applied Microbiology and Biotechnology,* 100(5), 2063-2072.

[30] Nair, L. S., and Laurencin, C. T. (2007). Biodegradable polymers as biomaterials. *Progress in Polymer Science*, *32*(8), 762-798.

[31] Wang, L., He, S., Wu, X., Liang, S., Mu, Z., Wei, J., & Wei, S. (2014). Polyetheretherketone/nano-fluorohydroxyapatite composite with antimicrobial activity and osseointegration properties. *Biomaterials*, *35*(25), 6758-6775.

[32] Lu, Z., Zhang, X., Li, Z., Wu, Z., Song, J., & Li, C. (2015). Composite copolymer hybrid silver nanoparticles: preparation and characterization of antibacterial activity and cytotoxicity. *Polymer Chemistry*, *6*(5), 772-779.

[33] Bilal, M., Rasheed, T., Iqbal, H. M. N., Hu, H., & Zhang, X. (2017). Silver nanoparticles: Biosynthesis and antimicrobial potentialities. *International Journal of Pharmacology*, *13*(7), 832-845.

[34] Bilal, M., Rasheed, T., Iqbal, H. M. N., Li, C., Hu, H., & Zhang, X. (2017). Development of silver nanoparticles loaded chitosan-alginate constructs with biomedical potentialities. *International journal of biological macromolecules*, *105*, 393-400.

[35] Rasheed, T., Bilal, M., Iqbal, H. M. N., & Li, C. (2017). Green biosynthesis of silver nanoparticles using leaves extract of Artemisia vulgaris and their potential biomedical applications. *Colloids and Surfaces B: Biointerfaces*, *158*, 408-415.

[36] Chang, W. S., & Chen, H. H. (2016). Physical properties of bacterial cellulose composites for wound dressings. *Food Hydrocolloids*, *53*, 75-83.

[37] Qiu, Y., Qiu, L., Cui, J., & Wei, Q. (2016). Bacterial cellulose and bacterial cellulose-vaccarin membranes for wound healing. *Materials Science and Engineering: C*, *59*, 303-309.

[38] Kiziltas, E. E., Kiziltas, A., Rhodes, K., Emanetoglu, N. W., Blumentritt, M., & Gardner, D. J. (2016). Electrically conductive nano graphite-filled bacterial cellulose composites. *Carbohydrate polymers*, *136*, 1144-1151.

[39] Barud, H. S., Tercjak, A., Gutierrez, J., Viali, W. R., Nunes, E. S., Ribeiro, S. J. L., Jafellici, M., Nalin M., and Marques, R. F. C. (2015). "Biocellulose-based flexible magnetic paper," *Journal of Applied Physics* 117(17), 17B734.

[40] Duarte, E. B., Bruna, S., Andrade, F. K., Brígida, A. I., Borges, M. F., Muniz, C. R., Filho, M. M. S., Morais, J. P. S., Feitosa, J. P. A., and Rosa, M. F. (2015). "Production of hydroxyapatite–bacterial cellulose nanocomposites from agroindustrial wastes," *Cellulose* 22, 3177-3187.

[41] Lin, C. W., Liang, S. S., Chen, S. W., & Lai, J. T. (2013). Sorption and transport properties of 2-acrylamido-2-methyl-1-propanesulfonic acid-grafted bacterial cellulose membranes for fuel cell application. *Journal of Power Sources*, *232*, 297-305.

[42] Ul-Islam, M., Khan, T., Khattak, W. A., & Park, J. K. (2013). Bacterial cellulose-MMTs nanoreinforced composite films: novel wound dressing material with antibacterial properties. *Cellulose*, *20*(2), 589-596.

[43] Feng, Y., Zhang, X., Shen, Y., Yoshino, K., & Feng, W. (2012). A mechanically strong, flexible and conductive film based on bacterial cellulose/graphene nanocomposite. *Carbohydrate Polymers*, *87*(1), 644-649.

[44] Shi, Z., Zang, S., Jiang, F., Huang, L., Lu, D., Ma, Y., & Yang, G. (2012). In situ nano-assembly of bacterial cellulose–polyaniline composites. *RSC Advances*, *2*(3), 1040-1046.

[45] Hu, W., Chen, S., Yang, Z., Liu, L., & Wang, H. (2011). Flexible electrically conductive nanocomposite membrane based on bacterial cellulose and polyaniline. *The Journal of physical chemistry B*, *115*(26), 8453-8457.

[46] Kim, J., Cai, Z., Lee, H. S., Choi, G. S., Lee, D. H., and Jo, C. (2011). "Preparation and characterization of a bacterial cellulose/chitosan composite for potential biomedical application," *Journal of Polymer Research* 18(4), 739-744.

[47] Zhu, H., Jia, S., Yang, H., Tang, W., Jia, Y., & Tan, Z. (2010). Characterization of bacteriostatic sausage casing: A composite of bacterial cellulose embedded with ε-polylysine. *Food Science and Biotechnology*, *19*(6), 1479-1484.